\documentclass[11pt,nofootinbib,showpacs]{revtex4}
\usepackage{amsfonts,amssymb,amsmath,graphicx}
\def\dac{\displaystyle\frac}

\def\[{\left[}
\def\]{\right]}
\def\({\left(}
\def\){\right)}

\newcommand{\diag}{\mathop{\rm diag}\nolimits}

\begin{document}

\baselineskip7mm
\title{New features of flat (4+1)-dimensional cosmological model with a perfect fluid in Gauss-Bonnet gravity}

\author{I.V. Kirnos}
\affiliation{Tomsk State University of Control Systems and
Radioelectronics, Tomsk, 634050 Russia}

\author{S.A. Pavluchenko}
\affiliation{Special Astrophysical Observatory, Russian Academy of
Sciences, Nizhnij Arkhyz, 369167 Russia}

\author{A.V. Toporensky}
\affiliation{Sternberg Astronomical Institute, Moscow State
University, Moscow, 119992 Russia}

\begin{abstract}
We investigated  a flat multidimensional cosmological model in
Gauss-Bonnet gravity in presence of a matter in form of perfect
fluid. We found analytically new stationary regimes (these results
are valid for arbitrary number of spatial dimensions) and studied
their stability by means of numerical recipes in $4+1$-dimensional
case. In the vicinity of the stationary regime we discovered
numerically another non-singular regime which appears to be
periodical. Finally, we demonstrated that the presence of matter
in form of a perfect fluid lifts some constraints on the dynamics
of the $4+1$-dimensional model which have been found earlier.
\end{abstract}


\maketitle

\section{Introduction}

Theories of gravity with the Gauss-Bonnet (GB) term as a correction
to the Einstein term in the action have been actively investigated
for more than 30 years~\cite{d4, raz1, Deruelle, raz2, deruelle2}.
This term appears as the first non-Einstein contribution in the
extension of the General Relativity (GR) known as the Lovelock
gravity~\cite{Lovelock} (the only theory that keeps the equations of
motion to be the 2nd order differential equations), and string
gravity~\cite{Metsaev}. In $3+1$ dimensions this term does not
contribute to the equations of motion, and it can be important
either in a combination with another fields (as it does in string
theory) or in theories containing scalar function of the
Gauss-Bonnet invariant (see, for example \cite{fR}; in this case we,
however, have higher-order equations of motion).

For higher number of spatial dimensions the Gauss-Bonnet term is
dynamically important. Moreover, in Lovelock gravity (in contrast
with string gravity) this term is the only allowed non-Einstein
contribution in ($4+1$) and ($5+1$) dimensions. This means that it
is possible to study regimes where this term is not a small
corrections to Einstein gravity, but equally important or even
dominant.

In the end of 80th some important results in multidimensional
cosmology with the Gauss-Bonnet term have been obtained including
the analog of Kasner solution in a pure Gauss-Bonnet gravity
\cite{Deruelle}. Further studies reveal interesting differences
between ($4+1$)-dimensional case (this is the lowest number of
dimensions for GB term to contribute) and higher-dimensional
cosmology. This includes differences in the form of power-law
solutions \cite{Deruelle, ivashchuk}, noncontinuity between vacuum
solutions and solutions with matter (in ($4+1$) flat case with
non-zero matter density $\rho$ the limit $\rho \to 0$ may not
coincide with vacuum solutions \cite{we2}). Some of the
differences can be extended to higher-order Lovelock corrections
(at least in a flat case~\cite{serg09}). Other example of
particularity of ($4+1$) case is severe fine-tuning needed for
smooth evolution from high density (where GB term dominates) to
low density (when Einstein gravity is restored) -- such evolution
requires three of four Hubble parameters to be equal with high
precision \cite{ind}. This result have been obtained numerically
in the theory which includes both GB and Einstein terms, so
abovementioned results on dynamics with matter \cite{we2} (founded
for a pure GB gravity) does not directly applicable to this case.
Nevertheless, as we shell see, nonzero matter density can alter
corresponding dynamics significantly.

The structure of the manuscript is as follows: first, we write
down equations of motion for the model considered. Then, we
analytically investigate stationary case and demonstrate exact
solutions found. Then we switch to numerical studies and confirm
by means of numerical recipes the reality of the previously found
solutions as well as describe new oscillationary regime. Also we
demonstrate (also via numerical methods) that the presence of
matter in form of a prefect fluid lifts some constraints on the
dynamics of the model considered. Finally, in the Conclusions we
summarize  the results found.

\section{Equations of motion}

We consider a flat anisotropic metric in (4+1)-dimensional
space-time. We are dealing with Einstein-Gauss-Bonnet gravity, and
non-vacuum space-time. Lagrangian of this theory have a form

$$
{\cal L} = R + \alpha{\cal L}_2 + {\cal L}_M,
$$

\noindent where $R$ is Ricci scalar, ${\cal L}_M$ is the
Lagrangian of matter fields and ${\cal L}_2$

\begin{equation}
{\cal L}_2 = R_{\mu \nu \alpha \beta} R^{\nu \mu \alpha \beta} - 4
R_{\mu \nu} R^{\mu \nu} + R^2 \label{lagr1}
\end{equation}

\noindent is the Gauss-Bonnet Lagrangian.

We are working in the flat background, so the metric we
considering has a form

\begin{equation}\label{metric}
g_{\mu\nu} = \diag\{ -1,a^2(t),b^2(t),c^2(t),d^2(t)\}.
\end{equation}

We use perfect fluid with the equation of state $p=w\rho$ as a
matter source; after varying action obtained from~(\ref{lagr1}) using the metric
above and perfect fluid as a matter field, one can obtain the
following equations of motion -- there are $n$ dynamical equations and a constraint
one. First dynamical equation has the form

\begin{eqnarray}
\begin{array}{l}
2(\dot H_b+H_b^2)+2(\dot H_c+H_c^2)  +2(\dot
H_d+H_d^2)+2H_bH_c+2H_bH_d+2H_cH_d + \\
\\+ 8\alpha\Big [ (\dot H_b+H_b^2)H_cH_d +(\dot H_c+H_c^2)H_bH_d
+(\dot H_d+H_d^2)H_bH_c\Big ] + \dac{16\pi
G}{c^4}w\varepsilon_0(abcd)^{-(1+w)}=0
\end{array}
 \label{dyn}
\end{eqnarray}

\noindent where as usual $H_i=\frac{\dot a_i}{a_i}$,
the rest of them could be obtained via cyclic index permutation. The constraint equation has a form

\begin{equation}
2H_aH_b+2H_aH_c+2H_aH_d+2H_bH_c+2H_bH_d+2H_cH_d+24\alpha
H_aH_bH_cH_d  = \dac{16\pi G}{c^4}\varepsilon_0(abcd)^{-(1+w)}.
 \label{constr}
 \end{equation}

\section{Stationary solutions}

The system of equations (\ref{dyn})--(\ref{constr}) has solutions of a particular type in which
all Hubble functions are constant (and, correspondingly, scale factors expand or
contract exponentially), so, we call them as stationary for brevity.
Anisotropic exponential solutions in Einstein-Gauss-Bonnet gravity
with matter were particularly studied in
\cite{kirnos-makarenko} under some assumptions. Here we study maximally anisotropic case:
$$g_{\mu\nu}=\{-1,e^{2H_1t},e^{2H_2t},\ldots,e^{2H_nt}\}.$$

For solution to be a stationary one
it is necessary that right-hand sides of field equations
$$\alpha_1 G^{(1)}_{\mu\nu}+\alpha_2 G^{(2)}_{\mu\nu}=\displaystyle\frac{8\pi
G}{c^4}T_{\mu\nu}$$ do not depend on time. This means that
we need
\begin{equation}\label{condition-exp}-(1+w)\sum_i
H_i=0,\end{equation} i. e. one of a two conditions should be satisfied:
\begin{enumerate}
\item $w=-1$ --- matter is a cosmological constant; or
\item $\sum_i H_i=0$ --- volume element is constant.
\end{enumerate}

As in this section we have some results for an arbitrary number of spatial
dimensions, we write down the  equation of motion for a flat
$(n+1)$-dimensional stationary case (i.e. neglecting all time derivatives):
\begin{equation}\label{eq00-exp}\alpha_1\sum_{i<j} H_i H_j +12\alpha_2\sum_{i<j<k<l}H_i H_j H_k
H_l=\displaystyle\frac{8\pi G}{c^4}\varepsilon_0,\end{equation}
\begin{equation}\label{eqjj-exp}\begin{array}{l}\displaystyle
\alpha_1\left(H_j\sum_i H_i-\sum_i H_i^2-\sum_{i<k}H_i
H_k\right)+2\alpha_2\left\{-2\sum_{i\neq j} H_i^2\sum_{\substack{k,l\neq i\\
k<l}}H_k H_l+2H_j\sum_{\substack{i,k\neq j\\ i\neq k}}H_i^2
H_k+\right.\\ \quad{}\left.\displaystyle+6H_j\sum_{\substack{i,k,l\neq j\\
i<k<l}}H_i H_k H_l- 6\sum_{i<k<l<m}H_i H_k H_l
H_m\right\}=\displaystyle\frac{8\pi
G}{c^4}w\varepsilon_0.\end{array}\end{equation}

For $w=-1$ we have investigated only $(4+1)$-dimensional space
($n=4$) with $\alpha_1=1$ ($\alpha\equiv\alpha_2$) and have obtained
the following solutions:
\begin{enumerate}
\item $$H_1=H_2=H,\qquad H_3=H_4=-\displaystyle\frac{1}{4\alpha H},$$ where $H$
satisfies condition $$\displaystyle\frac{8\pi
G}{c^4}\varepsilon_0=\displaystyle\frac{1-4\alpha H^2+16\alpha^2
H^4}{16\alpha^2 H^2}.$$
\item $$H_1\equiv H\mbox{ is arbitrary},\qquad
H_2=H_3=\displaystyle\frac{\zeta}{\sqrt{-4\alpha}},$$ $$\qquad
H_4=\displaystyle\frac{1-\zeta
H\sqrt{-4\alpha}}{\zeta\sqrt{-4\alpha}+4\alpha H},\qquad
\displaystyle\frac{8\pi
G}{c^4}\varepsilon_0=-\displaystyle\frac{3}{4\alpha},$$ where
$\zeta=\pm 1$.
\item $$H_1=H,\qquad H_2=\displaystyle\frac{z}{\alpha H},\qquad H_3=-\displaystyle\frac{1}{4\alpha
H},\qquad H_4=\displaystyle\frac{-16\alpha^2 H^4-1+16\alpha H^2
z-4z}{4\alpha H(-4\alpha H^2+8z+1)},$$ where
$$z=\displaystyle\frac{1-4\alpha H^2\pm\sqrt{5(4\alpha
H^2-1)^2+16\alpha H^2}}{8},$$ and $H$ satisfies condition
$$\displaystyle\frac{8\pi G}{c^4}\varepsilon_0=\displaystyle\frac{16\alpha^2 H^4-4\alpha
H^2+1}{16\alpha^2 H^2}.$$
\item $$H_3=-\displaystyle\frac{H_1+H_2}{2}\pm\sqrt{\displaystyle\frac{(H_1+H_2)^2(8\alpha H_1 H_2-3)+4H_1 H_2}{4(1+8\alpha H_1
H_2)}},$$ $$H_4=-\displaystyle\frac{-4\alpha H_1 H_2 (H_1^2+
H_2^2)+H_3(H_1+H_2)(1+8\alpha H_1 H_2)+H_1 H_2(1+4\alpha H_1
H_2)}{(1+8\alpha H_1 H_2)(H_1 +H_2 +H_3+12\alpha H_1 H_2 H_3)},$$
and $H_1,$ $H_2$ satisfies condition $$\displaystyle\frac{8\pi
G}{c^4}\varepsilon_0=\displaystyle\frac{4(H_1^2+H_2^2+H_1 H_2)\alpha
H_1 H_2}{1+8\alpha H_1 H_2}.$$
\item $$H_1=H_2=H,\qquad H_3=zH,$$ $$H_4=-\displaystyle\frac{H(288\alpha^2 H^4 z+8\alpha H^2 z-2z+52\alpha H^2-1)}{96\alpha^2 H^4 z -20\alpha H^2 z+
8\alpha H^2-z-2+1536z H^6\alpha^3+288\alpha^2 H^4},$$ where
$z=-1\pm\sqrt{\displaystyle\frac{8\alpha H^2-2}{8\alpha H^2+1}},$
and $H$ satisfies condition
$$\displaystyle\frac{8\pi G}{c^4}\varepsilon_0=\displaystyle\frac{12H^4\alpha}{1+8\alpha H^2}.$$
\item $$H_1=H_2=H_3=\pm\displaystyle\frac{1}{\sqrt{-4\alpha}},\qquad H_4\mbox{
is arbitrary,}\qquad \displaystyle\frac{8\pi
G}{c^4}\varepsilon_0=-\displaystyle\frac{3}{4\alpha}.$$
\end{enumerate}

For $\sum_i H_i=0$ a general solution was investigated for arbitrary number of
dimensions. If we express
$$\begin{array}{l}\displaystyle 24\sum_{i<j<k<l}H_iH_jH_kH_l=\sum_i H_i\sum_{j\neq
i}H_j\sum_{k\neq i,j}H_k\sum_{l\neq i,j,k}H_l=\vphantom{\displaystyle\frac 1 2}\\
\displaystyle\quad{}=\sum_i H_i\sum_{j\neq i}H_j\sum_{k\neq
i,j}H_k(-H_i-H_j-H_k)=-3\sum_i H_i^2\sum_{j\neq i}H_j\sum_{k\neq
i,j}H_k=\vphantom{\displaystyle\frac 1 2}\\
\displaystyle\quad{}=-3\sum_i H_i^2\sum_{j\neq
i}H_j(-H_i-H_j)=3\sum_i
H_i^3\sum_{j\neq i}H_j+3\sum_i H_i^2\sum_{j\neq i}H_j^2=\vphantom{\displaystyle\frac 1 2}\\
\displaystyle\quad{}=-3\sum_i H_i^4+3\sum_i H_i^2(\sum_j
H_j^2-H_i^2)=-6\sum_i H_i^4+3(\sum_i
H_i^2)^2\vphantom{\displaystyle\frac 1 2}\end{array}$$ and
similarly for other products of Hubble parameters, we get the
equations in the form

\begin{equation}\label{sigma}\begin{array}{l}-\alpha_1\sigma_2+3\alpha_2(\sigma_2^2-2\sigma_4)=2\varkappa_0,\vphantom{\displaystyle\frac 1 2}\\
-\alpha_1\sigma_2+2\alpha_2(\sigma_2^2-2\sigma_4)=(1+w)\varkappa_0,\end{array}\end{equation}
where $$\sigma_2\equiv\sum_i H_i^2,\quad \sigma_4\equiv\sum_i
H_i^4,\quad \varkappa_0\equiv\displaystyle\frac{8\pi
G}{c^4}\varepsilon_0.$$
If $\alpha_1\neq 0$ and $\alpha_2\neq 0$
then we obtain that the general solution is any set of $H_i,$ $i=1,\ldots
n,$ that satisfies the following conditions:
$$\begin{array}{l}\vphantom{\displaystyle\frac 1 2}\sum_i H_i=0,\\
\vphantom{\displaystyle\frac 1 2}\sum_i H_i^2=\displaystyle\frac{1-3w}{\alpha_1}\varkappa_0,\\
\vphantom{\displaystyle\frac 1 2}\sum_i
H_i^4=\displaystyle\frac{w-1}{2\alpha_2}\varkappa_0+\displaystyle\frac{(1-3w)^2}{2\alpha_1^2}\varkappa_0^2.\end{array}$$

From this we can obtain a particular case of \cite{kirnos-makarenko}
where $H_1=H_2=H_3,$ $H_4=H_5=\ldots=H_n.$

It is interesting to note that from equations (\ref{sigma}) in the
case $\alpha_1=0$ and $\varepsilon_0=0$ we have only one
condition to be satisfied

\begin{equation}\label{cond_vac}
\sum_i H_i^4=\displaystyle\frac 1 2 \left(\sum_i H_i^2\right)^2,
\end{equation}

and that is in agreement with \cite{ivashchuk}.
It can be also easily seen than the only case for a matter other than cosmological
constant allowing a stationary solution in the pure Gauss-Bonnet gravity
is of $w=1/3$ type.  In this case the condition is

\begin{equation}\label{cond_mat}
\left(\sum_i H_i^2\right)^2-2\sum_i
H_i^4=\displaystyle\frac{2\varkappa_0}{3\alpha_2}.
\end{equation}

One can easily see that (\ref{cond_mat}) turns into (\ref{cond_vac}) if $\varepsilon_0 = 0$. These two (\ref{cond_vac}, \ref{cond_mat}) are
the only special solutions of (\ref{sigma}) if $\alpha_1 = 0$.

\section{Numerical studies}

Before describing regimes let us make some notes regarding our strategy. First of all, just like in previous papers where
we numerically studied (4+1)- and (5+1)-dimensional Bianchi-I models~\mbox{\cite{we1, we2}}, we consider only models with positive
initial $\sum H_i$ (and in this section we again use dynamical $H_i = H_i(t)$ in contrast with the previous section where due to
parametrization used our Hubble function were constant). In (4+1) dimensions we specify 3 initial Hubble functions and calculate 4th using
the constraint equation~(\ref{constr}). This expression has a form

\begin{equation}
H_d^{(0)} = - \frac{ 2H_a^{(0)}H_b^{(0)} + 2H_a^{(0)}H_c^{(0)} +
2H_b^{(0)}H_c^{(0)}  - \rho_0 } {2H_a^{(0)} + 2H_b^{(0)} +
2H_c^{(0)} + 24\alpha H_a^{(0)}H_b^{(0)}H_c^{(0)}}, \label{4hubble}
\end{equation}

\noindent and here (and later) we use definition for $\rho_0$ (and later for $\rho$) to simplify notations: $\rho_0 = 16\pi G
\varepsilon_0 /c^4$.
Our goal is to study numerically the influence of non-zero matter density on the evolution
scenario. We fix 3 Hubble parameters and change $\rho$ and the equation of state
parameter $w$. In this approach two different possibilities exist depending on the
sign of denominator. If it is positive, we have no restriction on the initial value of matter
density $\rho$. If the denominator is negative, some maximal value of initial density exists (that maximal value satisfy
$\sum H_i^{(0)} = 0$).

Solving $\dot H_i = 0$ system with respect to ($w$, $\rho$) gives us values ($w_{st}$, $\rho_{st}$) for stationary regime:


\begin{equation}\label{wcr}\small
w_{st} = \dac{H_a^2+4 \alpha H_b H_a H_c^2+4 \alpha H_c H_b H_a^2+4 \alpha H_c H_a H_b^2+H_c^2+H_b H_a+H_c H_a+H_b^2+H_c H_b}
{H_b^2+H_c^2+H_a^2+H_c H_b+H_c H_a+H_b H_a+12 \alpha H_c H_b H_a^2+12 \alpha H_c H_a H_b^2+12 \alpha H_b H_a H_c^2},
\end{equation}\normalsize


\begin{equation}\label{rhocr}\small
\rho_{st} = -2 H_b^2-2 H_c^2-2 H_a^2-2 H_c H_b-2 H_c H_a-2 H_b H_a-24 \alpha H_c H_b H_a^2-24 \alpha H_c H_a H_b^2-24 \alpha
H_b H_a H_c^2,
\end{equation}\normalsize

\noindent and that regime corresponds to the one, described analytically in the previous section.

\subsection{Singular vacuum regimes}
We start with  listing of possible regimes  in (4+1)-dimensional
vacuum Bianchi-I model. There are two possible outcomes for the
past evolution -- standard and non-standard singularities, and
three for the future evolution -- non-standard singularity, Kasner
regime and recollapse. One can naively suppose that all six
possible combinations of ``past'' and ``future'' regimes are
occuring, but the reality is much more complicated. We summarized
the trajectories in Table.

\begin{table*}
{{\bf Table.} Possible trajectories for vacuum (4+1)-dimensional GR+GB model}\\
\begin{tabular}{c|c|c|c}
\hline
From & To & Design. & Avail. \\
\hline

Standard singularity & Kasner & I & + \\
 & Recollapse & II & + \\
 & Non-standard sing. & III & $-$\\
\hline

Non-standard. sing. & Kasner & IV & +\\
 & Recollapse & V & $-$ \\
 & Non-standard sing. & VI & + \\
\hline

\end{tabular}
\end{table*}

As it was noted in~\cite{ind}, type-I regime require fine-tuning
of initial Hubble parameters, namely, three of them should be
equal to each other. Recently~\cite{fulldesc} it was noted that
sometimes the equality of two Hubble parameters is enough to
achieve this type of regime.

In contradiction with~\cite{ind}, we failed to find type-III
transitions. Authors of paper~\cite{ind} claim they found them
(around 1.5\% of total number of trajectories), so keeping in mind
possible numerical errors as well as different criteria to stop
the integration process we can conclude that this regime is very
rare if exists at all (see also~\cite{fulldesc} for possible
reasons behind this).

\subsection{Regimes in presence of matter}

\begin{figure}
\includegraphics[width=0.7\textwidth,angle=-90,bb=74 35 565 720,clip]{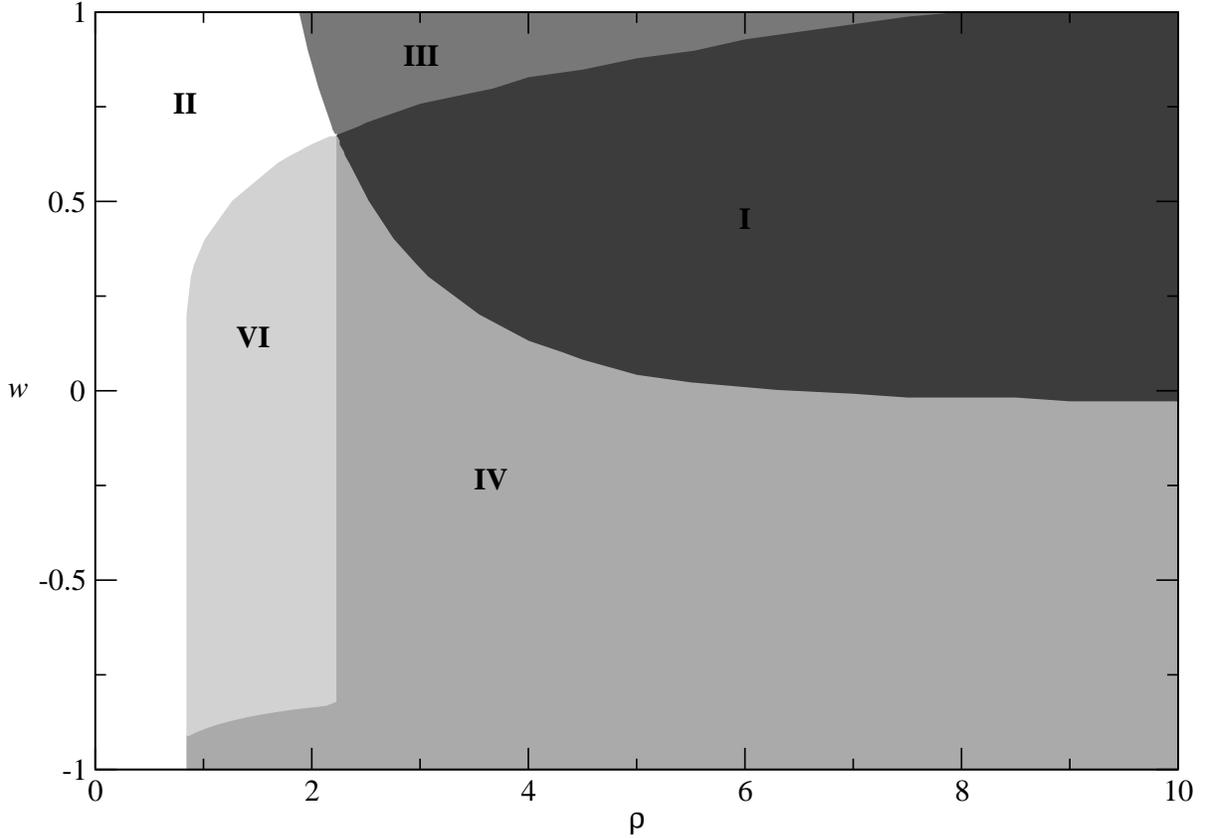}
\caption{An example trajectories map in $(\rho, w)$ coordinates
for II case with positive denominator. White region corresponds to
the type-II transition, different grey regions correspond to VI,
IV, III and I with increase of darkness of grey; they also denoted
in figure.}\label{2plus}
\end{figure}

\begin{figure}
\includegraphics[width=0.7\textwidth, angle=-90,bb=70 30 570 760, clip]{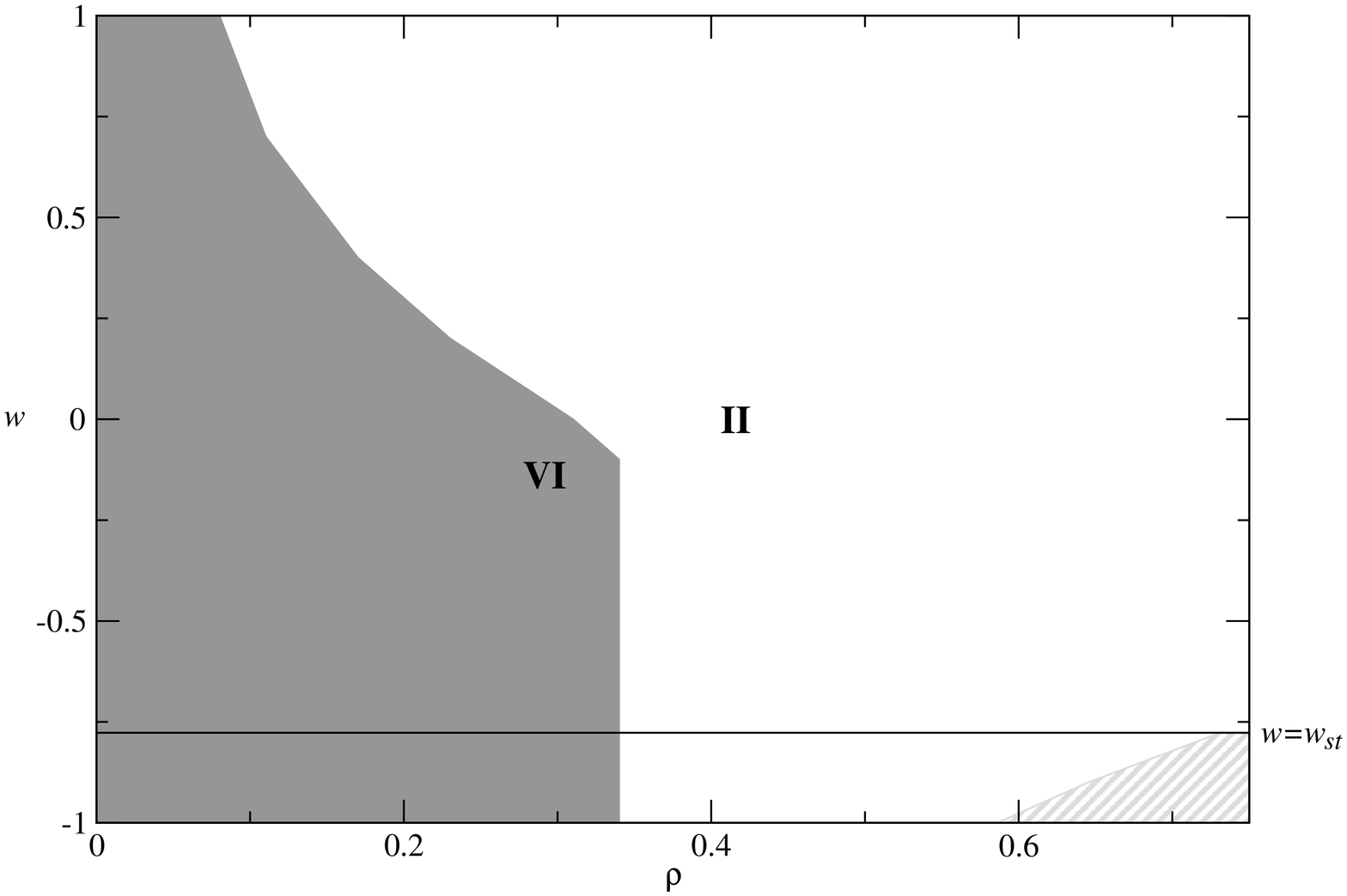}
\caption{An example trajectories map in $(\rho, w)$ coordinates
for VI case with negative denominator. Stationary regime occur at
$w=w_{st}$ and $\rho = \rho_{st}$, i.e. at the intersection of
$w=w_{st}$ with right $\rho$-boundary. Stroked region in the
right-bottom corner denotes periodical
trajectories.}\label{6minus}
\end{figure}

Since distinguishing of initial conditions with different signs of
denominator in Eq. (\ref{4hubble}) is important for us, we treat
these two cases separately. A typical transition diagram on the
$(\rho, w)$ plane for the case of positive denominator is given in
Fig. \ref{2plus} (to be precise, it corresponds to the vacuum
type-II case with positive denominator). White region corresponds
to the type-II behavior, different grey regions correspond to VI,
IV, III and I with increase of darkness of grey; they also denoted
in figure.

From this figure we can see that type III evolution (which we have
not seen in the vacuum case) and type I evolution (which occurs in
the vacuum case only under very specific initial conditions) both
appear as possible scenarios without any severe fine-tuning of
initial conditions. As a result, all 6 possible scenarios listed in
the Table 1, are possible when matter is taken into account.

In the transition diagram for the case of negative denominator we
have the stationary regime at the right limit of the diagram if
corresponding $w_{st}$ lies in the physical zone $w_{st}>-1$. In
Fig. \ref{6minus} as a way of example we present a typical example
for negative $w_{st}$ (that is type-VI case with negative
denominator). The situation at the limit line $\rho=\rho_{st}$ is
as follows.

\begin{figure}
\includegraphics[width=0.8\textwidth,bb=0 18 340 510,clip]{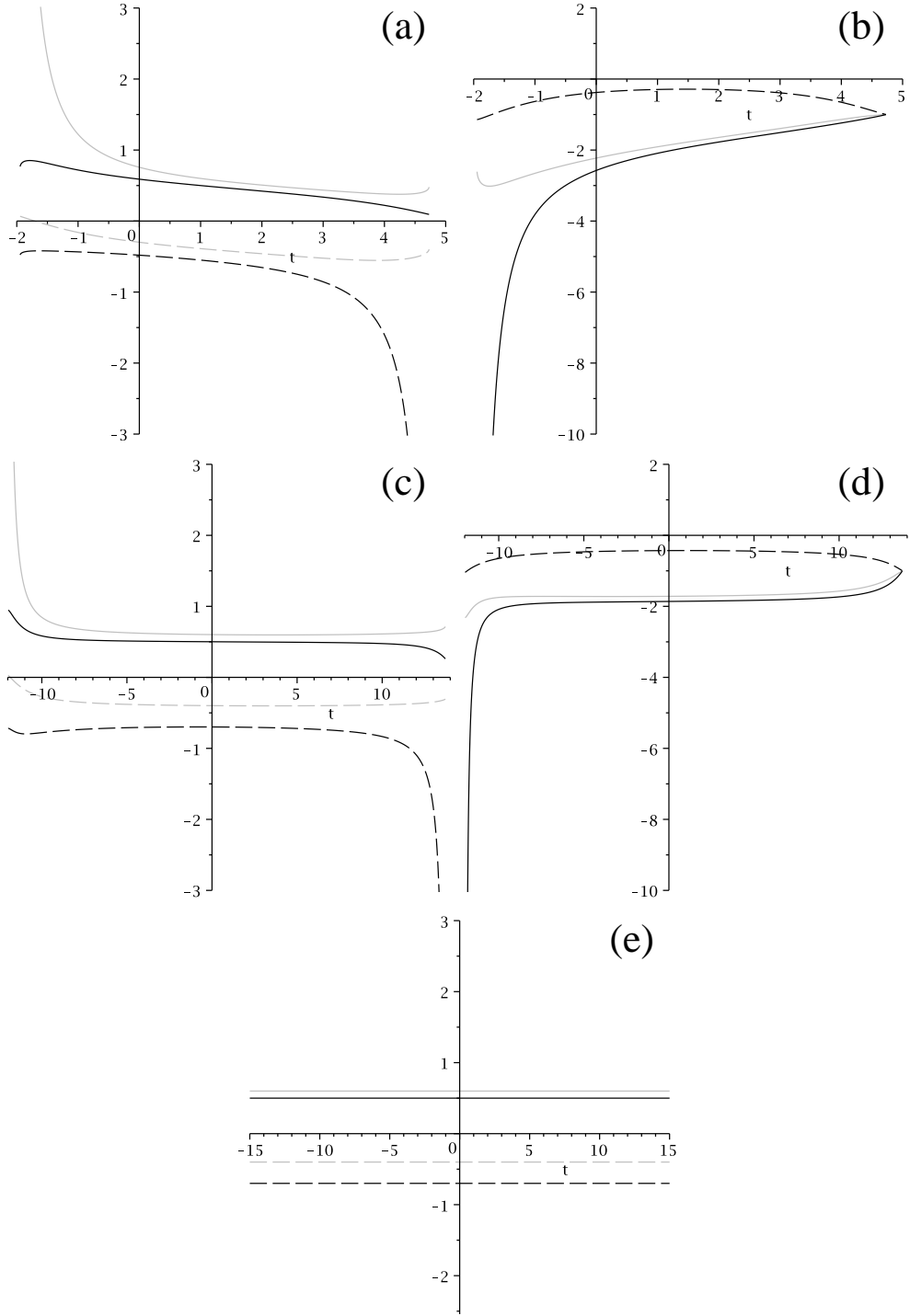}
\caption{The transition from standard type-II to stationary regime
with $w_{st} < 0$ and $w \to w_{st}+0$. Hubble functions in the
standard type-II regime are presented in (a); relative differences
$(H_i (t) - H_a(t))/H_a(t)$ in (b) demonstrate that the second
singularity (that corresponds to recollapse) is isotropic one.
With $w \to w_{st}+0$ the situation changes according to (c): the
``lifetime'' is increasing with ``central'' part looks pretty
flat; the second singularity remains isotropic one (d). Finally,
when $\rho = \rho_{st}$ and $w = w_{st}$, the regime becomes
stationary  one (e).}\label{fig1}
\end{figure}

If $w > w_{st}$ and $w \to w_{st}+0$ the ``lifetime'' of the
type-II regime is increasing and asymptotically reaches infinity
when $w = w_{st}$. In Fig. \ref{fig1} we presented the transition
in question: Fig. \ref{fig1}(a) represents Hubble functions versus
time for standard type-II regime (standard singularity to
recollapse); one can see in Fig. \ref{fig1}(b), where we plot
$(H_i (t) - H_a(t))/H_a(t)$ that the second singularity (that
corresponds to recollapse) is isotropic one. With $w \to w_{st}+0$
(and $\rho \to \rho_{st}$) the situation changes according to Fig.
\ref{fig1}(c); second singularity is still isotropic (Fig.
\ref{fig1}(d)). Finally, when $\rho = \rho_{st}$ and $w = w_{st}$,
the stationary regime is reached (Fig. \ref{fig1}(e)).

\begin{figure}
\includegraphics[width=0.8\textwidth,bb=0 18 340 510,clip]{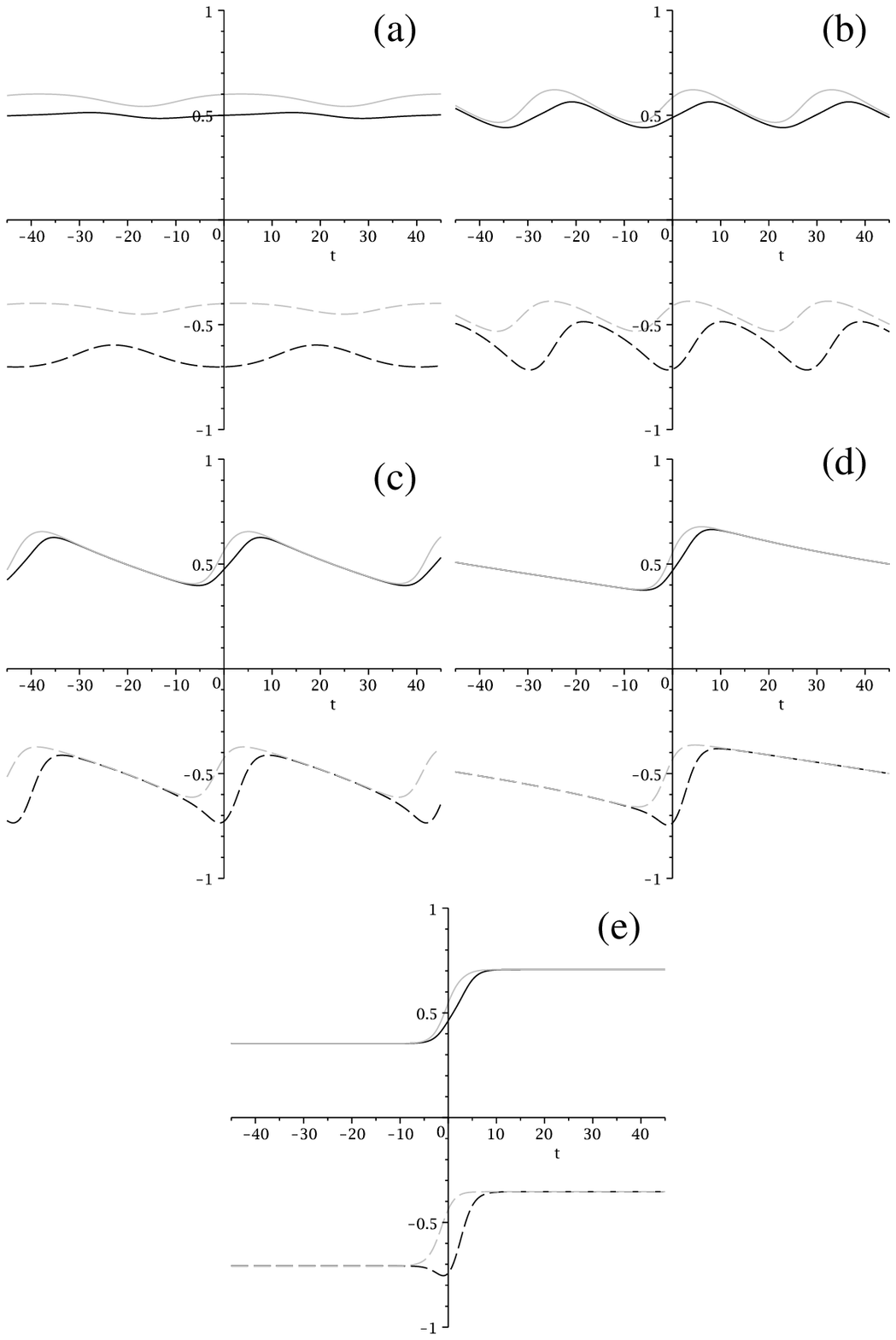}
\caption{The transition from standard type-II to stationary regime
with $w_{st} < 0$ and $w \to w_{st}-0$. We used the same $\rho
\approx \rho_{st}$ for all panels (a)--(e) and decreased the
equation of state from $w \approx w_{st}-0$ in panel (a) down to
$w=-1$ in panel (e). See text for details.}\label{fig2}
\end{figure}

If $w < w_{st}$ we have quite different behavior. Namely, it
becomes periodical; in Fig. \ref{fig2} we presented it. All
through Fig. \ref{fig2} we used the same value for density $\rho
\approx \rho_{st}$ and decreased equation of state $w$ from $w
\approx w_{st}-0$ in Fig. \ref{fig2}(a) down to $w=-1$ in Fig.
\ref{fig2}(e). One can see that with decrease of the equation of
state the period and amplitude of oscillations are first
increasing first, and then smoothly turn into quasi-stationary
regime at $w=-1$. This means that small homogeneous perturbation
of stationary regime does not necessary destroys nonsingular
behavior, but turn it to periodic oscillations near the stationary
point. If $\rho$ is less enough than $\rho_{st}$  then the
described above periodical regime is not triggered and we have
singular regime for any $w$.

\begin{figure}
\includegraphics[width=0.7\textwidth,angle=-90,bb=74 35 565 760,clip]{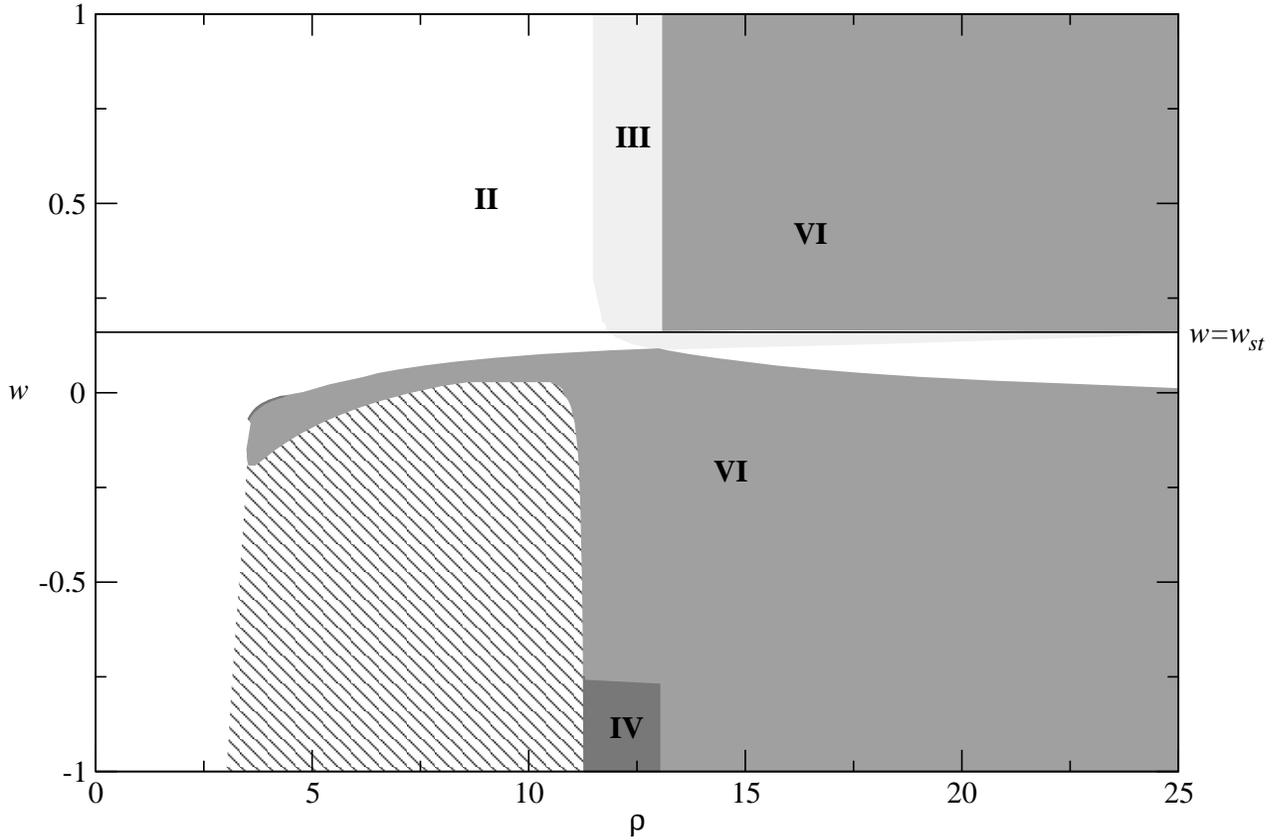}
\caption{An example map of transitions in $(\rho, w)$ coordinates for $1/3 > w_{st} > 0$ case. White corresponds to type-II transitions;
stroked white -- periodical trajectories as described in $w_{st} < 0$ subsection; light grey -- type-III trajectories; grey -- type-VI
and dark grey -- type-IV.}\label{life2}
\end{figure}

In case of $w_{st}>0$ the situation is different from described
above. There are no periodic regimes in the vicinity of the
stationary point. Any small perturbations of initial conditions
needed for this solution ultimately leads to singular behavior. An
example for regime transitions diagram for this case is plotted in
Fig. \ref{life2}.
In general, we have not seen any periodic oscillations numerically with $w>0$.

\section{Conclusions}
We have considered dynamics of a flat $(4+1)$-dimensional
anisotropic Universe filled with an ordinary matter in
Gauss-Bonnet gravity. Regarding singular regimes, we have founded
that any of possible 6 different regimes (with 2 possible initial
points in standard or nonstandard singularities, and 3 possible
future outcomes -- recollapse, nonstandard singularity or
low-curvature Kasner solution) can be realized without severe
fine-tuning of initial conditions. This means that some
constraints of $(4+1)$-dimensional dynamics in Gauss-Bonnet
gravity found for vacuum regimes are lifter when matter is taken
into account. Full description of all transition in both vacuum
and matter cases is given in~\cite{fulldesc}.

We have found also two nonsingular regimes -- a stationary one with
constant values of Hubble parameters and zero volume expansion rate
(so, $\sum H_i=0$) which have been found analytically for an
arbitrary number of dimensions  and oscillatory regime founded numerically in the vicinity
of the stationary regime for $(4+1)$-dimensional case. In our
numerical studies we have seen this latter regime only for $w<0$.

\section*{Acknowledgements}
This work was partially supported by RFBR grant 08-02-00923.

\end{document}